%Paper: hep-th/9211052
%From: Franco.Ferrari@LS-WESS.PHYSIK.UNI-MUENCHEN.DBP.DE
%Date: Wed, 11 Nov 92 23:34:46 +0100

\font\subtit=cmr12
\font\name=cmr8
\input harvmac
\def\LMU#1#2#3{\TITLE{LMU-TPW \number\yearltd-#1}{#2}{#3}}
\def\TITLE#1#2#3{\nopagenumbers\abstractfont\hsize=\hstitle\rightline{#1}%
\vskip 1in\centerline{\subtit #2}%
\vskip 1pt
\centerline{\subtit #3}\abstractfont\vskip .5in\pageno=0}%
\LMU{13} {MULTIVALUED FIELDS ON THE COMPLEX PLANE}
{AND CONFORMAL FIELD THEORIES}
\centerline{F. F{\name ERRARI}\foot{\name Work supported
by the Consiglio Nazionale Ricerche, P.le A. Moro 7, Roma, Italy}}\smallskip
\centerline{\it Sektion Physik der Universit\"at M\"unchen}\smallskip
\centerline{\it Theresienstr. 37, 8000 M\"unchen 2}\smallskip
\centerline{\it Fed. Rep. Germany}
\vskip 2cm
\centerline{ABSTRACT}
\vskip 1truecm
{\narrower In this paper a class of conformal field theories with nonabelian
and discrete
group of symmetry is investigated. These theories are realized in terms of
free scalar fields starting from the simple
$b-c$ systems and scalar fields on algebraic curves.
The Knizhnik-Zamolodchikov equations
for the conformal blocks can be explicitly solved.
Besides of the fact that one obtains
in this way an entire class of theories in which the operators
obey a nonstandard statistics,
these systems are interesting in exploring the connection between statistics
and curved space-times, at least in the two dimensional case.}
\Date{August 1992}
%\draft
\newsec{INTRODUCTION}
\vskip 1cm
In this paper we investigate the connections between
conformal field theories on the complex plane
and field theories on algebraic curves.
These connections were first explored in \ref\zam{
Al. B. Zamolodchikov, {\it Nucl. Phys.} {\bf B285 [FS19]}
(1987) 481.} in the case of hyperelliptic curves and then in
\ref\knirev{V. G. Knizhnik, {\it Sov. Phys. Usp.} {\bf 32}(11)
(1989) 945.}, \ref\brzn{
M. A. Bershadsky and A. O. Radul, {\it Int. Jour. Math.
Phys.} {\bf A2} (1987) 165.},
\ref\brrmp{
M. A. Bershadsky and A. Radul,
{\it Phys. Lett.} {\bf 193B} (1987), 21.} in the more general
case of curves with abelian group of monodromy.
Other examples of these techniques, in which the monodromy group is abelian,
are given in
\ref\others{
L. Dixon, D. Friedan, E. Martinec, S. Shenker,
{\it Nucl. Phys.} {\bf B282} (1987), 13;
S. Hamidi, C. Vafa, {\it Nucl. Phys.} {\bf B279} (1987), 465;
J. J. Atick, A. Sen, {\it Nucl. Phys.} {\bf B286} (1987), 189;
L. Bonora, M. Matone, F. Toppan and K. Wu, {\it Phys. Lett.}
{\bf 224B} (1989) 115; Nucl. Phys. {\bf B334} (1990) 717;
E. Guadagnini, M. Martellini, M. Mintchev, {\it Jour. Math. Phys.}
{\bf 31} (1990), 1226.}, \ref\raina{
A. K. Raina, {\it Helv. Phys. Acta} {\bf 63} (1990), 694.}.\smallskip
Here we study the simplest class of curves with nonabelian group of
monodromy.
They can be viewed as multivalued mappings from the complex sphere to a
Riemann surface having a discrete group of authomorphisms $D_m$.
Alternatively they can be viewed as cyclic coverings of hyperelliptic
curves.
The case $m=3$ was shortly treated in
\ref\ferdthree{
F. Ferrari, {\it Phys. Lett} {\bf 277B} (1992), 423.}.
\smallskip
In general, the construction of the amplitudes of a theory with
nontrivial monodromy properties require the solution of a Riemann
monodromy problem (RMP) and of the related Schlesinger equations
\ref\sjm{
M. Sato, T. Miwa and M. Jimbo. Holonomic quantum fields
(Kyoto U.P. Kyoto), part I; 14 (1978) p. 223; II: 15 (1979) p. 201;
III: 15 (1979) p. 577; IV: 15 (1979) p. 871; V; 16 (1980) p.
531.}, \ref\chud{D. V. Chudnowski in: C. Bardos, D. Bessis (eds.),
Bifurcation Phenomena in Mathematical
Physics and Related Topics, D. Reidel Publishing Company 1980.}.
Even if we are able to solve the RMP, still remains the problem of
determining what combinations of the solutions enter in the amplitudes,
in such a way that the physical properties of locality, associativity
and so on are preserved
\ref\bk{B. Blok and S. Yankielowicz, {\it Nucl. Phys.}
{\bf B321} (1989) 327.
B. Blok and S. Yankielowicz, {\it Phys. Lett.}
{\bf 226B} (1989) 279.}.
In the case in which the monodromy group coincides with the monodromy
group $G$ of a known algebraic curve, there is the possibility of
simplifications, since the most general function exchanging its branches
according to $G$, can be constructed using the techniques of algebraic
geometry \ref\ferstrone{
F. Ferrari, {\it Int. Jour. Mod. Phys.} {\bf A5} (1990), 2799.}.\smallskip
This is the case for example in which the monodromy group $G$ describes
the class of algebraic curves with discrete group of symmetry
$D_m$. For these curves we can in fact construct a finite set of
functions (and more in general $\lambda-$differentials)
$F_k(z)$, $k=0,\ldots,2m-1$, characterized by all possible
monodromy properties that are compatible with the monodromy group $G$.
We show that the elements of this set are rationally independent, i.e.
the ratio of two of them is not a singlevalued function and that
all the other multivalued functions are linear combinations of
the $F_k(z)$'s.
Moreover, our set of functions satisfies partial
differential equations similar to the equations of parallel transport
for the conformal blocks of
\ref\kohno{
T. Kohno, {\it Nagoya Math. J.} {\bf 92}
(1983), 21; {\it Invent. Math.} {\bf 82} (1985), 57.},
\ref\tk{
A. Tsuchiya, Y. Kanie, {\it Lett. Math. Phys.} {\bf 13}
(1987), 303.}.
Finally, following refs. \sjm, we show that it is possible to express the
multivalued functions $F_k(z)$ in terms of free fields and twist fields.
Therefore,
starting from the $F_k(z)'s$, we are able to construct conformal blocks,
whose monodromy properties correspond to the monodromy group
$G$.\smallskip
It is difficult to associate a conformal field theory defined on the complex
plane to these conformal blocks. However they are surely tightly related
to the $b-c$ systems on the algebraic curve $\Sigma_g$ with $D_m$ group
of symmetry, as we will see.\smallskip
The method presented here is interesting because it allows the
construction of conformal blocks with nontrivial monodromy properties,
provided the underlying monodromy group is that of a known algebraic
curve. Moreover, the twist fields turns out to be anyons, exchanged in
the conformal blocks according to a non abelian braid group statistics.
Unlike the usual anyons realized starting from a nonabelian Chern$-$Simons
field theory \ref\bw{B. Blok, X. G. Wen,
{\it Nucl. Phys.} {\bf b374} (1992), 615.}, the exchange relations between
the twist fields become nonabelian due to the presence of the group of
authomorphisms $D_m$ of the algebraic curve.
The statistics of the twist fields has been studied in a separate
publication \ref\ffbgs{F. Ferrari, Preprint LMU-TPW 92-24}.
Finally we provide a nice interpretation of the twist fields as
electrostatic charges induced by the topology of the algebraic
curve.\smallskip
The disadvantage of our approach is that we are not able to prove that
the conformal blocks satisfy a Riemann monodromy problem.
For example, the relationships between the branch points are
too complicated in order to derive the Schlesinger equations.
Moreover, our method can be surely extended to other classes
of curves but not, we believe, to the most general cases
where, appearently, there seems to be obstructions in the construction
of some of the
functions $F_k(z)$ which satisfy the requirements
given in Section 2.\smallskip
The material contained in this paper is organized as follows.
In Section 2 we find the conditions for which a $\lambda$ differential
on a general algebraic curve can be represented as a ratio of conformal
blocks containing free fields and twist fields. The general form of the
twist fields is given.
Starting from Section 3 we restrict ourselves to
the class of $D_m$ symmetric curves.
We construct a basis of $\lambda$ differentials
satisfying the conditions of Section 2.
They are rationally independent and exhibit all possible monodromy behaviors
at the branch points compatible with the monodromy group
of the algebraic curve.
Moreover, all other meromorphic $\lambda$ differentials are linear
combinations of them.
In Section 4 the full $n$-point functions of the $b-c$
systems on the $D_m$ symmetric curves
are computed.
The $n$-point functions turn out to be superpositions of the solutions of the
conformal blocks defined in Section 3.
In Section 5 it is shown that the $b-c$ systems on an
algebraic curve with $D_m$ group of symmetry contain multivalued
operators with fractional ghost charges.
These twist fields simulate the presence of the branch points in the
amplitudes and are primary fields. The appearance of primary fields in the
amplitudes of the $b-c$ systems is explained in terms of electrostatics
in Section 6.
In Section 7 the form of the
twist fields is explicitly given in terms of free fields
using bosonization and the method introduced in Section 2.
We prove that, apart from zero modes, the two-point function of the
$b-c$ systems on an algebraic curve can be seen as
conformal field theories. Each conformal field theory is characterized by
particular monodromy properties at the branch points of the algebraic curve.
The conformal blocks satisfy differential equations of the kind of the
Knizhnik-Zamolodchikov equations \ref\kz{
V. G. Knizhnik, A. B. Zamolodchikov, {\it Nucl. Phys.}
{\bf B247} (1984), 83.}, \kohno, \tk.
Finally the exchange relations between the twist fields are derived
showing that they
satisfy a nonabelian braid group statistics
\ref\froa{
J. Fr\"ohlich, Statistics of fields, The Yang-Baxter equation
and the theory
of knots and links, in: Nonperturbative quantum field theory, eds. G.
t'Hooft et al. (Plenum, New York, 1988).},
\ref\mack{
G. Mack, V. Schomerus, {\it Nucl. Phys.} {\bf B370} (1992),
185.}.
\vskip 1cm
\newsec{MONODROMY PROPERTIES AND TWIST FIELDS}
\vskip 1cm
Let us consider a classical field $B(z)dz^\lambda$, $\lambda$ integer or
half$-$integer, satisfying a Fermi statistics, analytic in $z$ and
taking its values on an affine algebraic curve $\Sigma_g$ defined by the
vanishing of a Weierstrass polynomial $F(z,y)$:
\eqn\weierstrass{F(z,y)=P_n(z)y^n+\ldots+P_0(z)=0}
Each affine algebraic curve is equivalent, apart from conformal
transformations, to a closed and orientable Riemann surface. The genus
$g$ of the Riemann surface is given by the Riemann$-$Hurwitz formula
\ref\gh{
H. Griffiths, J. Harris, Wiley Interscience, New York
1978.},
which we will not discuss here.
The $P_i(z)$'s, $i=0,\ldots,n$, are polynomials in the complex variable
$z\in {\rm\bf CP}_1$, ${\rm\bf CP}_1$ denoting the Riemann sphere.
Here it is useful to regard the sphere as a compactified complex plane
${\rm\bf C}\cup\{\infty\}$, covered by the two open sets $U_1$ and $U_2$
which contain the points $0$ and $\infty$ respectively.
$z$ is the local coordinate in $U_1$ and $z'$ in $U_2$. At the
intersections of these two sets $z'=1/z$. Solving eq.
\weierstrass\ in $y$, we get a multivalued function $y(z)$ with $n$
branches, denoted here by $y^{(l)}(z)$, $l=0,\ldots,n-1$. As a
consequence, the complex field $B(z)dz^\lambda$ becomes multivalued when
transported along a closed small path encircling the branch points of
$y(z)$:
\eqn\multivaluedness{B^{(l)}(z)dz^\lambda\equiv
B(z,y^{(l)}(z))dz^\lambda}
On an algebraic curve, $dz^\lambda$ represents a true $\lambda$
differential with zero and poles \ferstrone.
The degree of its divisor is $2\lambda(g-1)$. Therefore we can consider
$B^{(l)}(z)$ in eq. \multivaluedness\ as a function multiplied by the
$\lambda$ differential $dz^\lambda$.
Let us suppose that $B^{(l)}(z)$ has zeros $z_i$ and
poles $p_j$, $i,j=1,\ldots,N$ of multiplicities $\nu_i(l_i)$ and $\mu_j(l_j)$
respectively.
The zeros and poles occur only for certain values of the branch $l$ of
$B^{(l)}(z)$ and therefore the multiplicities $\nu(l_i)$ and
$\mu_j(l_j)$ should also depend on the branch index.
Now we associate to $B^{(l)}(z)dz^\lambda$ another
$1-\lambda$ differential defined as follows:
\eqn\cdiff{C^{(l)}(z)dz^{1-\lambda}={dz^{1-\lambda}\over B^{(l)}(z)}}
At this point we investigate the conditions under which the tensor
$G(z,w)dz^\lambda dw^{1-\lambda}=B^{(l)}(z)C^{(l')}(w){dz^\lambda
dw^{1-\lambda}\over z-w}$ in the two independent complex variables $z$ and $w$
can be written in terms of conformal blocks. To
this purpose, we introduce free fields $\tilde b(z)dz^\lambda$ and
$\tilde c(z)dz^{1-\lambda}$ on $\Sigma_g$, which are however singlevalued
in the variable $z$.
Since they do not depend on $y(z)$, their expansion is the usual Laurent
series of the genus zero case.
The fields $\tilde b(z)$ and $\tilde c(z)$ are fermions or ghosts
according to the values of $\lambda$.
Moreover we introduce multivalued
``twist fields"
$V(z_i)$ and $V(p_j)$ with the following multivalued operator product
expansions (OPE):
\eqn\twists{\eqalign{\tilde b(z) V^{(l_i)}
(z_i)\sim&(z-z_i)^{\nu_i(l_i)}:\tilde b(z)
V^{(l_i)}(z_i):+\ldots\cr
\tilde b(z) V^{(l_j)}(p_j)\sim&(z-p_j)^{-\mu_j(l_j)}:\tilde b(z)
V^{(l_j)}(p_j):+\ldots\cr
\tilde c(z) V^{(l_i)}(z_i)\sim&(z-z_i)^{\nu_i(l_i)}:\tilde c(z)
V^{(l_i)}(z_i):+\ldots\cr
\tilde c(z) V^{(l_j)}(p_j)\sim&(z-p_j)^{-\mu_j(l_j)}:\tilde c(z)
V^{(l_j)}(p_j):+\ldots\cr}}
Apart from zero modes, which we ignore for the moment, we express the
tensor $G(z,w)dz^\lambda dw^{1-\lambda}$ in the form:
\eqn\main{{B_k^{(l)}(z)
C_k^{(l')}(w)\over z-w}dz^\lambda dw^{1-\lambda}
={<0|\tilde b(z)\tilde c(w)
\prod\limits_{i=1}^N V(z_i)
\prod\limits_{j=1}^N V(p_j)|0>\over
<0|\prod\limits_{i=1}^{N}V_k(z_i)
\prod\limits_{j=1}^NV(p_j)|0>}}
$<0|$ being the usual vacuum at genus zero. For the twist fields $V(z_i)$
and $V(p_j)$ we can try the simple ansatz of
\ferdthree:
\eqn\twistfields{\eqalign{V_k^{(l)}(z_i)=&{\rm exp}\left[i\oint_{C_{z_i}}dt
\partial_t{\rm
log}\left[C^{(l)}(t)\right]\varphi(t)\right]\cr
V_k^{(l)}(p_j)=&{\rm exp}\left[i\oint_{C_{p_j}}dt
\partial_t{\rm
log}\left[B^{(l)}(t)\right]\varphi(t)\right]\cr}}
after using bosonization:
\eqn\bosone{\tilde b(z)\sim{\rm e}^{-i\varphi(z)}\qquad\qquad\qquad
\tilde c(z)\sim{\rm e}^{i\varphi(z)}}
\eqn\bostwo{<0|\varphi(z)\varphi(w)|0>=-{\rm log}(z-w)}
The multivaluedness of the twist fields, caused by the fact that
the zeros and poles
of $B^{(l)}(z)$ and $C^{(l)}(z)$ occur only for certain values
of the branches, implies
that they are nonlocal operators in the most general
case, as equation \twistfields\ shows.
Moreover, since the OPE's with the free fields turns out to be multivalued,
the right hand side (rhs) in
\main\ is also multivalued in $z$ and $w$. Consistently with the
left hand side (lhs), the branches in $z$ and $w$ of the rhs should be $l$ and
$l'$ respectively.
We remember here another similar example in which the presence of
nonabelian groups
of symmetries introduce nonlocal fields in the amplitudes, namely
the solitonic sectors of scalar field theories with discrete
group of symmetries discussed in \froa, \ref\fr2{
J. Fr\"ohlich, P. A. Marchetti, {\it Comm. Math. Phys.}
{\bf 112} (1987), 343.}.
Exploiting eq. \bostwo, we evaluate the
OPE's between the twist fields and the free
fields as in the genus zero case.
More OPE's are not needed to evaluate eq. \main.
Proceeding as in \ferdthree\ we can compute the rhs of
\main\ obtaining the following result:
\eqn\ope{B^{(l)}(z)C^{(l')}(w)
={\rm exp}\left[-
\left(\sum\limits_{i=1}^N\oint_{C_{z_i}}+
\sum\limits_{i=1}^N\oint_{C_{p_j}}\right)
\partial_t{\rm log}C(t){\rm
log}\left({t-w\over t-z}\right)
\right]}
Here we have used the fact that, by definition,
${\rm log} B(z)=-{\rm log}C(z)$. Eq. \ope\ can be rewritten as follows:
\eqn\opesimpl{B^{(l)}(z)C^{(l')}(w)dz^\lambda dw^{1-\lambda}=
{\rm exp}\left[-\oint_C\partial_t{\rm log}C(t){\rm log}\left({t-w\over
t-z}\right)\right]}
$C$ being a contour surrounding all poles and zeros of $C(z)$.
Unfortunately it is impossible to apply the theorem of residues in
\opesimpl. The function in the integrand is in fact multivalued inside
the contour $C$ and, in general, also on the contour itself. For this
reason we additionally require that all the branch points of
$B^{(l)}(z)$ are included in the set of points $z_i$ and
$p_j$.
This is a reasonable request in view of our applications, since in
conformal field theories on an algebraic curve the physical zeros and poles
in the amplitudes are given by the branch points of the algebraic curve
(see Section 4).
Under the above requirement, the integrand in \opesimpl\ becomes
onevalued on the contour $C$ because it surrounds all the branch points
of $B^{(l)}(t)$ and $C^{(l)}(t)$.
Moreover, since we are on the compact sphere ${\rm\bf CP}_1$, we can
deform the contour $C$ in such a way that only the other two
singularities of the integrand are included, namely the points $t=z$ and
$t=w$.
The integration by parts in the exponent of eq. \opesimpl\ is then made
possible and yields:
\eqn\intparts{\oint_C\partial_t{\rm log}C(t){\rm log}\left({t-w\over
t-z}\right)= \oint_{C_w+C_z}{\rm log} C(t)\left({1\over t-w}-{1\over
t-w}\right)}
$C_w+C_z$ describes a simple contour equivalent to $C$ containing the
points $w$ and $z$. The integrand of the lhs of eq. \intparts\ is now
onevalued inside and on the contour $C_w+C_z$, so that we can easily
compute its residue:
\eqn\residue{\oint_{C_w+C_z}{\rm log}C(t)\left({dt\over t-w}-{dt\over
t-z}\right)={\rm log}C(w)-{\rm log}C(z)}
Substituting eq. \residue\ in the rhs of eq. \opesimpl\ we obtain an
identity, proving that eq. \main\ makes sense if all the ramification points
of $B^{(l)}(z)$ are included in the set $z_i$ and $p_j$.
\vskip 1cm
\newsec{CONFORMAL BLOCKS FOR THE $b-c$ SYSTEMS ON AN ALGEBRAIC CURVE}
\vskip 1cm
At this point we specify a class of Riemann surfaces $\Sigma_g$ of genus $g$
associated to the Weierstrass polynomial
\eqn\curve{y^{2m}-2q(z)y^m+q^2(z)-p(z)=0}
$q(z)$ and $p(z)$ are polynomials in the variable $z$.
The genus $g$ is given in
Appendix A in terms of the degrees $mr$ and $2r'$ of $q(z)$ and $p(z)$
respectively.
The algebraic curve $y(z)$ has $2m$ branches denoted by $y^{(l)}(z)$,
$l=0,\ldots,2m-1$, that are exchanged at the branch
points $\alpha_i$ and $\beta_j$ as shown in Appendix A. $i$ and $j$ label
the number of the independent roots of the equations
\eqn\branchp{q^2(z)-p(z)=0\qquad\qquad\qquad p(z)=0} respectively.
The first equation has $N_\alpha={\rm max}(2mr,2r')$ solutions
$\alpha_i$ while the second equation has $N_\beta= 2r'$ solutions $\beta_j$.
The integers $r$ and $r'$ are fixed in such a way that
the point at infinity is not a branch point.
This is not an essential limitation and it is introduced only in order to
keep the notations as simple as possible.
\smallskip
Eq. \curve\ is invariant under a $D_m$ group of symmetry, generated
by the transformations:
\eqn\dmsymm{(z,y)\rightarrow(z,\epsilon y)\qquad\qquad{\rm and}\qquad\qquad
y^m-q(z)\rightarrow-y^m+q(z)}
where $\epsilon^m=1$. The local monodromy group
contains $D_m$ as a subgroup.
It is possible to view $\Sigma_g$ as a $Z_m$ cyclic covering of an
hyperelliptic curve $H_g$ of genus $g'=r'+1$ and branch points $\beta_j$.
The multivaluedness at the branch points $\alpha_i$ is then related
to the $Z_m$ branched covering of $H_g$.\smallskip
We start constructing a basis $B_k(z)$, $0\le k\le 2m-1$,
of $2m$ rationally independent
functions on $\Sigma_g$ such that all other functions
are linear combinations of them, the coefficients entering the linear
combination being at most
singlevalued functions of $z$.
Two functions are said to be rationally independent if their ratio
is not a singlevalued function on ${\rm\bf CP}_1$. A basis of that kind
is for example given by $B_k(z)=[y(z)]^k$, $0\le k\le 2m-1$.
However, the elements of this basis do not satisfy in general the
requirement to have all their ramification points included in their
divisor.
Therefore we seek for a basis $B_k^{(l)}(z)$ with the following leading
order expansions at the branch points:
\eqn\localmonodromy{\eqalign{B_k^{(l)}(z)
\sim(z-\alpha_i)^{-q_{k,\alpha_i}(l)}+\ldots\cr
 B_k^{(l)}(z)
\sim(z-\beta_j)^{-q_{k,\beta_j}(l)}+\ldots\cr}}
Transporting the functions $B_k^{(l)}(z)$ around a branch point on a
closed path, one obtaines the phases ${\rm exp}(-2\pi
iq_{k,\alpha_i}(l))$, ${\rm exp}(-2\pi
iq_{k,\beta_j}(l))$ that depend on the initial branch $l$ of the function
and on the index $k$ characterizing the rationally independent
functions.
The $q_{k,\alpha_j}(l)$ and $q_{k,\beta_j}(l)$ must be rational numbers for
some values of $l$, otherwise there is no multivaluedness at all.
In principle, in order to find the $B_k^{(l)}(z)$,
one needs to solve a Riemann monodromy problem and the related
Schlesinger equations
\sjm, \chud, \bk.
However, this is not so simple and the boundary conditions of the
Schlesinger equations are not known.
Fortunately we can rely on a theorem of algebraic geometry stating that
a general function on an algebraic curve, therefore also a function
satisfying eqs. \localmonodromy, should be a rational function in $y(z)$
and $z$.
The construction of a function with a nontrivial behavior at the branch
points of the kind \localmonodromy\ can be done using techniques of
algebraic geometry. The parameters $q_{k,\alpha_i}(l)$ and
$q_{k,\beta_j}(l)$, however, are still defined only up to integers.
For example one can multiply them with singlevalued functions whose zeros
lie at the branch points.
This freedom is fixed by the physical properties that the
correlation functions of the conformal
field theories should satisfy, for example associativity, locality and
statistics of the fields.\smallskip
In this paper we choose a particularly simple conformal field theory, the
$b-c$ systems \ref\fms{
D. Friedan, E. Martinec, S. Shenker, {\it Nucl. Phys.}
{\bf B271} (1986), 93.} with spin $\lambda$ and action:
\eqn\bc{S=\int_{\Sigma_g}d^2zb\bar\partial c+{\rm c.c.}}
$b(z)dz^\lambda$ and $c(z)dz^{1-\lambda}$ are now fields on $\Sigma_g$
and consequently they are multivalued fields in $z$ in the sense of eq.
\multivaluedness.
For each value of $\lambda$, the physical requirements mentioned above
are dictated by the fermionic statistics of the $b-c$ systems.
In other words, their correlation functions should have simple poles
whenever the
coordinates of one field $b$ and one field $c$ coincide and simple zeros
in the case in which the coordinates of two fields $b$ or two fields $c$
coincide
\ref\vv{
E. Verlinde and H. Verlinde, {\it Nucl. Phys. } {\bf B288}
(1987) 357.}, \ref\bi{
M. Bonini and R, Iengo, {\it Int. Jour. Mod. Phys.} {\bf A3}
(1988) 841.}.
It is easy to check that, as a consequence, the parameters
$q_{k,\alpha_i}(l)$ and $q_{k,\beta_j}(l)$
must depend also on $\lambda$.
Therefore it is convenient to introduce two different basis
$B_k^{(l)}(z)$ and $C_k^{(l)}(z)$ for
the fields $b$ and $c$ respectively.
Finally, the freedom of multiplying the basis with a singlevalued function
with zeros and poles at the branch points will be exploited in such a way that
the correlation functions of the $b-c$ systems on $\Sigma_g$ can be expanded
in the simplest way in the elements of the basis.\smallskip
First of all we consider the case $\lambda=0$.
The following $2m$ functions $F_k(z)$ are an example of a basis
satisfying
the above requirements and those of Section 2:
\eqn\bbasiszero{
\eqalign{F_k(z)=&y^k(z)\qquad\qquad\qquad\qquad\qquad 0\le k\le m-1\cr
=&y^{2m-1-k}(z)\sqrt{p(z)}\qquad\qquad\quad m\le k\le 2m-1\cr}}
It is easy to check that the functions $F_k(z)$ are rationally
independent and that they have the behavior \localmonodromy\ at the
branch points with nontrivial rational values of $q_{k,\alpha_i}(l)$ and
$q_{k,\beta_i}(l)$.
\smallskip
Now we will prove that any rational function $R(z,y(z))$ of
$z$ and $y(z)$ is a linear combination of the functions $F_k(z)$ of the
kind:
\eqn\linear{R(z,y^{(l)}(z))=\sum\limits_kc_k(z)F_k^{(l)}(z)}
where the coefficients $c_k(z)$ are singlevalued in $z$.
Eq. \linear\ is certainly true if $R(z,y(z))$ is a sum of monomials of $z$
and $y(z)$.
In fact, from eq. \curve\ we have $y^m(z)=q(z)\pm\sqrt{p(z)}$.
Therefore monomials
containing powers in $y(z)$ greater than $m-1$ are still
expressible in terms of the basis \linear.
At this point we have only to consider the rational functions of the kind
$$R(z,y(z))={1\over \sum\limits_k c_k(z)F_k(z)}$$
A simple consequence of eq. \curve\ is the following equation:
\eqn\dependence{
\prod\limits_{l=0}^{m-1}\left(\sum\limits_kc_k(z)\epsilon^{kl}F_k(z)
\right)\left[R(z,y(z))\right]^{-1}=Q(z)\sqrt{p(z)}+P(z)}
$Q(z)$ and $P(z)$ being singlevalued in $z$.
Therefore
\eqn\rational{
R(z,y(z))={\left(Q(z)\sqrt{p(z)}-P(z)\right)\prod\limits_{l=0}^{m-1}
\left(\sum\limits_kc_k(z)\epsilon^{kl}F_k(z)\right)\over
Q^2(z)p(z)-P^2(z)}}
which is again of the kind \linear.
Thus we have shown that the functions $F_k(z)$ are $2m$ multivalued,
rationally independent functions and that all other functions, the solutions
the RMP included, are
linear superpositions of them.\smallskip
The case of general $\lambda$ is solved as follows.
As can be seen from the divisors written in
Appendix A, the $\lambda$-differential
\eqn\bbzero{
B_0(z)dz^\lambda={dz^\lambda\over [y(z)]^{\lambda(m-1)}[p(z)]^{-\lambda
\over 2}}}
has neither poles nor zeros at the branch points.
Therefore, multiplying $B_0(z)dz^\lambda$ with the functions $F_k(z)$ of eq.
\bbasiszero, we get $2m$ $\lambda$-differentials $B_k(z)dz^\lambda$
with all the possible independent behaviors at the branch points.
The final result is:
\eqn\basislambda{\eqalign{B_k^{(l)}(z)dz^\lambda=&
\left(y^{(l)}(z)\right)^{mq_{k,\alpha_i}}\left(p(z)
\right)^{q_{k,\beta_j}}dz^\lambda\cr
C_k^{(l)}(z)dz^{1-\lambda}=&
\left(y^{(l)}(z)\right)^{-mq_{k,\alpha_i}}\left(p(z)
\right)^{-q_{k,\beta_j}}dz^{1-\lambda}\cr}}
where the `` charges " $q_{k,\alpha_i}$ and $q_{k,\beta_j}$ are defined by:
\eqn\qka{q_{k,\alpha_i}={[k]_m+\lambda(1-m)\over m}\qquad\qquad\qquad
[k_m]=[k+m]_m=k}
and
\eqn\qkb{\eqalign{q_{k,\beta_j}=&-{\lambda\over 2}\qquad\qquad\qquad
k=0,\ldots,m-1\cr
=&{1-\lambda\over 2}\qquad\qquad\qquad k=m,\ldots,2m-1\cr}}
The significance of charges of the parameters $q_{k,\alpha_i}$ and
$q_{k,\beta_j}$ will be clarified below (see also \ferdthree,
\ref\ferbos{
F. Ferrari, {\it Jour. Math. Phys.} {\bf 32} (8), (1991) 2186.}).
It is easy to show that the elements in the basis \basislambda\ are
rationally
independent and that the functions $B_k(z)$, $C_k(z)$ are linear combinations
with rational coefficients of the $F_k(z)$'s.
The leading order behavior of $B_k^{(l)}(z)$ and $C_k^{(l)}(z)$ at the
branch points is again of the form given in eq. \localmonodromy.
The parameters $q_{k,\alpha_i}(l)$ and $q_{k,\beta_j}(l)$ are given by:
\eqn\qkla{\eqalign{q_{k,\alpha_i}(l)=&0\qquad\qquad\qquad0\le l\le m-1\cr
=& q_{k,\alpha_i}\qquad\qquad\qquad m\le l\le 2m-1\cr}}
and
\eqn\qklb{
q_{k,\beta_j}(l)=q_{k,\beta_j}\qquad\qquad\qquad 0\le l\le 2m-1}
\vskip 1cm
\newsec{THE $n$-POINT FUNCTIONS OF FREE FIELD THEORIES
ON A $D_m$ SYMMETRIC ALGEBRAIC CURVE}
\vskip 1cm
In this section we derive the correlation functions of the $b-c$ systems
showing that they are superpositions of the basis
given in eq. \basislambda.
The $N_b=(2\lambda-1)(g-1)$ zero modes
$\Omega_{1,\lambda}(z)dz^\lambda,\ldots,
\Omega_{N_b,\lambda}(z)dz^\lambda$ are computed in Appendix A in terms of
the basis \basislambda.
In the Appendix we have however exploited a different notation to number
the zero modes introducing a double index $i_k,k$.
The index $k$ labels the sector of zero modes having the same behavior at
the branch points of the $\lambda$-differential $B_k(z)dz^\lambda$,
while $i_k$ labels the zero modes inside a given sector.
This notation stress the fact that the zero modes are contructed in terms
of the basis \basislambda. Here, however, it
complicates the expressions of the
correlation functions and therefore will not be used.\smallskip
When $\lambda>1$, the following meromorphic tensor with a single
pole in $z=w$ will be necessary:
\eqn\kl{K^{(ll')}_\lambda(z,w)dz^\lambda dw^{1-\lambda}={1\over 2m}
{dz^\lambda dw^{1-\lambda}\over z-w}\sum\limits_{k=0}^{2m-1}
B_k^{(l)}(z)C_k^{(l')}(w)}
If $\lambda=1$ we need instead a differential of the third kind $\omega_{ab}
(z)dz$ with two simple poles in $z=a$ and $z=b$ with residue $+1$ and
$-1$ respectively:
\eqn\thirdkind{
\omega_{a_{(l')}b_{(l'')}}^{(l)}(z)dz=K_{\lambda=1}^{(ll')}(z,a)dz-
K_{\lambda=1}^{(ll'')}(z,b)dz}
The pole in $z=a$ is active only if $l=l'$. Analogously there is a
divergence in $z=b$ only if $l=l''$.
The zero modes, the tensors \kl\ and the differentials of the third kind
\thirdkind\ are derived using the formalism developed in ref. \ferstrone.
At this point we are ready to compute the $n$-point functions exploiting
the method of fermionic construction \ref\vv{
E. Verlinde and H. Verlinde, {\it Nucl. Phys. } {\bf B288}
(1987) 357.}, \ref\bi{
M. Bonini and R, Iengo, {\it Int. Jour. Mod. Phys.} {\bf A3}
(1988) 841.}.
For $\lambda>1$ the $n$-point functions are ratios of the following
correlators \bi:
$$
<\prod\limits_{s=1}^Mb^{(l_s)}(z_\rho)\prod\limits_{t=1}^Nc^{(l'_t)}
(w_t)>=$$
\eqn\bclambda{{\rm det}\left|\matrix{\Omega_{ 1,\lambda}^{(l_1)}(z_1)&\ldots&
\Omega_{ N_b,\lambda}^{(l_1)}(z_1)&K_\lambda^{(l_1l'_1)}(z_1,w_1)&
\ldots&K_\lambda^{(l_1l_N')}(z_1,w_N)\cr
\vdots&\ddots&\vdots&\vdots&\ddots&\vdots\cr
\Omega_{ 1,\lambda}^{(l_M)}(z_M)&\ldots&
\Omega_{ N_b,\lambda}^{(l_M)}(z_M)&K_\lambda^{(l_Ml'_1)}(z_M,w_1)&
\ldots&K_\lambda^{(l_Ml_N')}(z_M,w_N)\cr}\right |}
where $M-N=(2\lambda-1)(g-1)=N_b$.
The tensor $K_\lambda^{(ll')}(z,w)$ has spurious poles in the limit
$w\rightarrow\infty$. However one can show as in \ferstrone\ and \bi\
that these poles do not contribute to the determinant \bclambda.
For $\lambda=1$ we have an analogous equation:
$$<\prod\limits_{i=1}^Nb^{(l_i)}(z_i)\prod\limits_{j=1}^Mc^{(l'_j)}(w_j)>=
$$
\eqn\bcone{{\rm det} \left|\matrix{
\omega^{(l_1)}(z_1)_{w_2w_1}&\ldots&\omega^{(l_1)}(z_1)_{w_Mw_1}\enskip
\Omega_{ 1,1}^{(l_1)}(z_1)&
\ldots&\Omega_{ g,1}^{(l_1)}(z_1)\cr
\vdots&\ddots&\vdots&\ddots&\vdots\cr
\omega^{(l_N)}
(z_N)_{w_2w_1}&\ldots&\omega^{(l_N)}(z_N)_{w_Mw_1}\enskip
\Omega_{ 1,1}^{(l_N)}
(z_N)&
\ldots&\Omega_{ g,1}^{(l_N)}(z_N)\cr}\right |}
where $N-M=g-1$.
In order to simplify the notations we have omitted in the rhs of eq. \bcone\
the indices of the branches for the variables $w_1,\ldots,w_M$.
In the next section we will mainly use the two point functions of the $b-c$
systems:
\eqn\bctwopoint{
G_\lambda^{(ll')}(z,w)={<b^{(l)}(z)c^{(l')}(w)\prod\limits_{s=1}^{N_b}
b^{(l_s)}(z_s)\prod\limits_{r=1}^{N_c}c(w_r)>\over
det\left|\Omega_{\bar s,\lambda}(z_t)\right|}}
{}From eq. \bclambda\
it turns out that the above propagator has the following form:
$$G_\lambda(z,w)dz^\lambda dw^{1-\lambda}=K_\lambda^{(ll')}(z,w)dz^\lambda
dw^{1-\lambda}+$$
\eqn\splitting{\sum\limits_{s=1}^{N_b}(-1)^sK_\lambda^{(ll')}(z_s,w)
{<b^{(l_1)}(z_1)\ldots b^{(l_{s-1})}(z_{s-1})b^{(l)}(z)b^{(l_{s+1})}(z_{s+1})
\ldots b^{(l_{N_b})}(z_{N_b})>\over
<b^{(l_1)}(z_1)\ldots b^{(l_{N_b})}(z_{N_b})>}}
where
\eqn\determinant{
<b^{(l_1)}(z_1)\ldots b^{(l_{N_b})}(z_{N_b})>={\rm det}\left|
\Omega_{\bar i,\lambda}^{(l_j)}(z_j)\right|}
Eq. \splitting\ and the form of $K_\lambda^{(ll')}(z,w)$ given in eq.
\kl\
show that the propagator \bctwopoint\ is just a superposition
of the elements of the basis
\basislambda.
An analogous equation can be written when $\lambda=1$.\smallskip
Before concluding this section, we prove that also the correlation functions
of the scalar fields
\eqn\scalar{S[X]=\int_{\Sigma_g}d^2z\partial_zX\partial_{\bar z}X}
can be expressed as linear combinations of the elements of the basis
\basislambda.
The correlator $<\partial X X>$ is a differential of the third kind that
coincides with the propagator of the $b-c$ systems with $\lambda=1$
up to zero modes in $z$:
$$<\partial_z X(z,\bar z)\left[X(w,\bar w)-X(w',\bar w')\right]>=$$
\eqn\xcorr{\Re
\left[<b(z)c(w)\prod\limits_{i=1}^gb(z_i)c(w')>+{\rm zero}\enskip
{\rm modes}\right]+(w\rightarrow w')}
where the symbol $\Re[T]$ means that the real part of the tensor $T$ is
taken.
The correlation function $<\partial X \partial X>$ can be obtained
deriving eq. \xcorr\ in $w$ and $\bar w$.
The derivation in $w$ of the correlation function of the $b-c$ systems
\splitting\
is quite simple. The variable $w$ appears only in the tensor
$K_{\lambda=1}^{(l_sl')}(z_s,w)$ and $K_{\lambda=1}^{(ll')}(z,w)$.
The latter tensors are linear combinations of the basis
$C_k(w)dw^{1-\lambda}$ and can be easily differentiated using eqs.
(B.3), (B.4) of Appendix B.
\vskip 1cm
\newsec{CONFORMAL FIELD THEORIES WITH ${\rm D}_m$ GROUP OF SYMMETRY}
\vskip 1cm
In this Section we prove that the $b-c$ systems on an algebraic curve
are, apart from zero modes,
a conformal field theory, in the sense that they contain primary
fields concentrated at the branch points.
To this purpose we study the vacuum expectation values (vev's) of the ghost
current $J(z)=:b(z)c(z):$ and of the energy momentum tensor at the branch
points.
These vev's can be computed starting from the two point functions
\splitting.
We start considering
the vev of the ghost current, which is given by:
\eqn\normord{
<J^{(l)}(z)>=\lim_{\scriptstyle z\to w\atop l=l'}\left[G_\lambda^{(ll')}
(z,w)dz^\lambda dw^{1-\lambda}-{dz^\lambda dw^{1-\lambda}\over z-w}\right]
}
{}From eq. \splitting\ it is clear that the divergences at the branch points
are generated only by the term $K_\lambda^{(ll')}(z,w)dz^\lambda
dw^{1-\lambda}$.
The other terms forming the propagator are in fact zero modes in $z$
and the poles in the variable $w$ occur only at the locations of the zero
modes $z_s$ or in $z=\infty$.
Therefore, inserting eq. \kl\ in eq. \normord\ we get:
\eqn\gcurr{<J_z^{(l)}(z)>dz={1\over 2m}\sum\limits_{k=0}^{2m-1}
\partial_z{\rm log}C_k^{(l)}(z)dz}
It is possible to regard the differential
\eqn\jkdiff{J_k^{(l)}(z)dz=\partial_z{\rm log}\left[C_k^{(l)}(z)\right]dz}
as the vev of the current associated
to  the ghost number conservation in a given sector $k$, i. e. in the sector
in which the fields have the same monodromy properties of $B_k(z)dz^\lambda$
and $C_k(z)dz^{1-\lambda}$.
The leading order of eq. \gcurr\ at the branch points confirms eqs.
\qkla-\qklb:
\eqn\ja{
\eqalign{<J^{(l)}(z)>\sim&{\rm reg.}\enskip{\rm terms}\qquad\qquad\qquad
\qquad\qquad 0\le l\le m-1\cr
\sim&{1\over 2m}\sum\limits_{k=0}^{2m-1}\sum\limits_{i=1}^{N_\alpha}
{q_{k,\alpha_i}\over z-\alpha_i}\qquad\qquad\qquad m\le l\le 2m-1\cr}}
and
\eqn\jb{<J^{(l)}(z)>\sim {1\over 2m}
\sum\limits_{k=0}^{2m-1}\sum_{j=1}^{N_\beta}
{q_{k,\beta_j}\over z-\beta_j}}
Now we compute also the vev of $<T(z)>$ at
the branch points in the first order approximation.
As before, the only contribution comes from the
tensor $K^{(ll')}_\lambda(z,w)dz^\lambda dw^{1-\lambda}$ as we will show
immediately. The proof is a slight generalization of a simple argument
given in \knirev\ in the case of $Z_m$ symmetric curves.\smallskip
On the algebraic curve
$\Sigma_g$ the fields $b$ and $c$ are singlevalued. Therefore, in the
proper system of coordinates, the energy momentum tensor must be regular.
The proper coordinate near a branch point $\alpha_i$
is defined as follows:
\eqn\locuna{\eqalign{t=z&\qquad\qquad\qquad 0\le l\le m-1\cr
t^m=z-\alpha_i&\qquad\qquad\qquad m\le l\le 2m-1\cr}}
At the points $\beta_j$ the local uniformizer becomes instead:
\eqn\locunb{t^2=z-\beta_j\qquad\qquad\qquad 0\le l\le 2m-1}
Since the vev of the energy momentum tensor is not a tensor, a change of
coordinates like that given in eqs. \locuna\ and \locunb\ yields an extra term
which is nothing but a partial derivative:
\eqn\emt{<T(t)>=\left ({dz\over dt}\right)^2<T(z)>-{c_\lambda\over 6}\left [
{d^3z/dt^3\over dz/dt}-{3\over 2}\left ({d^2z/dt^2\over dz/dt}\right)^2
\right]}
where $c_\lambda=(6\lambda^2-6\lambda-1)$.
The Schwarzian derivative appearing in the rhs of equation \emt\
gives poles of the second order at the branch points.
In order to eliminate these singularities from $<T(t)>$,
the correlator $<T(z)>$
in eq. \emt\ should have the same singularities but with opposite signs, i.e.:
\eqn\emta{
\eqalign{<T(z)>=& {\rm reg.}\enskip{\rm terms}\qquad\qquad\qquad\qquad\qquad
0\le l\le m-1\cr
=& {1\over (z-\alpha_i)^2}{c_\lambda\over 12}\left ({1\over m^2}-1\right)
\qquad\qquad\qquad 0\le l\le 2m-1\cr}}
and at the branch points $\beta_j$:
\eqn\emtb{
<T(z)>={1\over (z-\beta_j)^2}{c_\lambda\over 16}\qquad\qquad\qquad\qquad
0\le l\le 2m-1}
Now we compute the correlator $<T(z)>$ explicitly.
We apply the following formula given in \vv:
\eqn\nord{<T^{(l)}(z)>=\lim_{\scriptstyle z\to w\atop l=l'}\left[-\lambda
\partial_wG_\lambda^{(ll')}(z,w)+(1-\lambda)\partial_z G_\lambda^{(ll')}
(z,w)-{1\over (z-w)^2}\right]}
After inserting in eq. \nord\ the tensor $K_\lambda^{(ll')}
(z,w)$ instead of the entire propagator the result is:
\eqn\tcorr{<T^{(l)}(z)>={1\over 2m}\sum\limits_{k=0}^{2m-1}\left[\left(
\lambda-{1\over 2}\right)\left ({d^2C_k^{(l)}(z)/dz^2\over C_k^{(l)}(z)}
\right)-(\lambda-1)\left({dC_k^{(l)}(z)/dz\over C_k^{(l)}(z)}\right)^2
\right]}
In the first order approximation at the branch points eq. \tcorr\ becomes:
\eqn\ta{
\eqalign{<T^{(l)}(z)>\sim&{\rm reg.}\enskip{\rm terms}\qquad\qquad\qquad
0\le l\le m-1\cr
\sim& {1\over 2m}\sum\limits_{k=0}^{2m-1}\left[{1\over 2}q_{k,\alpha_i}^2+
\left(\lambda-{1\over 2}\right)q_{k,\alpha_i}\right]{1\over(z-\alpha_i)^2}
\qquad\qquad m\le l\le 2m-1\cr}}
\eqn\tb{<T^{(l)}(z)>\sim{1\over 2m}\sum_{k=0}^{2m-1}\left[{1\over 2}
q_{k,\beta_j}^2+\left(\lambda-{1\over 2}\right)q_{k,\beta_j}\right]
{1\over (z-\beta_j)^2}\qquad\qquad\qquad 0\le l\le 2m-1}
Summing over $k$ in eqs. \ta\ and \tb\ we obtain exactly
eqs. \emta\ and \emtb.
Concluding, we have shown that the amplitudes of the $b-c$ systems
on an algebraic curve with
$D_m$ group
of symmetry contain primary fields with charges and conformal
dimensions given by eqs. \ja-\jb\ and \nord-\tcorr\ respectively.
In Section  6 we interprete these primary fields as
twist fields simulating the presence of the branch points in the
correlation functions.
\vskip 1cm
\newsec{ON THE GEOMETRICAL MEANING OF THE TWIST FIELDS AND THEIR ELECTROSTATIC
INTERPRETATION}
\vskip 1cm
In this Section we notice an important point coming from
the previous analysis.
The $b-c$ systems are singlevalued on $\Sigma_g$, so that the energy
momentum tensor has no singularities at the branch points in the proper
coordinates \locuna\ and \locunb.
Instead, the poles of the ghost current remain \foot{We thank J. Sobczyk
for having pointed out this fact}.
They can be explained as a topological effect induced by the fact
that we are considering a theory on a curved space$-$time. Already Wheeler
pointed out that topology is equivalent to charge on some manifolds.
For example in figure 1 the total effect of the potential lines
is to simulate a positive charge inside the left hole
and a negative charge inside the right one.
In our case, we have a more complicated surface, similar to that of figure 1
but with many
handles. Therefore it is natural to interprete the
poles of the ghost current at the branch points
as virtual (and fractional)
ghost charges generated by the nontrivial topology of the world$-$sheet.
On that point see also ref. \ferdthree.\smallskip
Now we explain this phenomenon in a somewhat heuristic way
using electrostatic considerations.
We consider the $b-c$ theory on $\Sigma_g$ as a multivalued field theory
on ${\rm\bf CP}_1$:
\eqn\bccpone{S^{(l)}(b,c)=\int_{{\rm\bf CP}_1}d^2zb^{(l)}(z)\bar\partial
c^{(l)}(z) +{\rm c.c.}}
As we showed in Section 3, the fields $b^{(l)}(z)$ and $c^{(l)}(z)$ can
be expanded in the basis \basislambda:
\eqn\modeexpansion{\eqalign{b^{(l)}(z)=\sum\limits_{k=0}^{2m-1}B_k^{(l)}(z)
\tilde b_k(z)\cr
c^{(l)}(z)=\sum\limits_{k=0}^{2m-1}C_k^{(l)}(z)
\tilde c_k(z)\cr}}
where $\tilde b_k(z)$ and $\tilde c_k(z)$ are singlevalued fields on
${\rm\bf CP}_1$, interacting only if $k=k'$ as the usual $b-c$ systems
on the sphere.
This expansion is valid only locally, i.e. away from the branch points
and from the point at infinity. At these points one should use the local
uniformizer and the coordinate $z'=1/z$ respectively.
Eq. \modeexpansion\ defines an operator formalism on $\Sigma_g$ and,
substituting eq. \modeexpansion\ in eq. \bccpone, we get:
\eqn\bcsplitted{S^{(l)}(b,c)=\int_{{\rm\bf CP}_1}
d^2z\sum\limits_{k=0}^{2m-1} \tilde b_k(z)\bar\partial \tilde c_k(z)+
\int_{{\rm\bf CP}_1}
d^2z\sum\limits_{k=0}^{2m-1} \bar\partial{\rm log} (C^{(l)}(z)\tilde
b_k(z) \tilde c_k(z)}
Now the fields are considered as operators, so that everywhere a
normal ordering should be understood.
In eq. \bcsplitted\ the multivaluedness of the action
is in the second term of the rhs.
At this point we can bosonize the action \bcsplitted\ using the formulas
\bosone\ and \bostwo\ for each field $\tilde b_k(z)$ and $\tilde
c_k(z)$.
As a consequence, after an integration by part, eq. \bcsplitted\
becomes:
\eqn\bosonizedaction{S^{(l)}(\varphi_k(z)=
\int_{{\rm\bf CP}_1}
d^2z\sum\limits_{k=0}^{2m-1}\left [\partial
\varphi_k\bar\partial\varphi_k+ R_{z\bar z}(z,\bar z)\varphi_k\right]+
\int_{{\rm\bf CP}_1}
d^2z\sum\limits_{k=0}^{2m-1}\partial\bar\partial {\rm
log}\left|C^{(l)}_k(z)\right|^2\varphi_k}
The term in $R_{z\bar z}(z,\bar z)$ comes from the usual
bosonization of the $b-c$ systems on the sphere and it is given by the
distribution:
\eqn\cpcurvature{R_{z\bar z}(z,\bar z)=(1-2\lambda)\delta^{(2)}(z,\infty)}
The third term in the rhs of eq. \bosonizedaction\ represents instead the
additional curvature requested by the fact that we treating a $b-c$ system
on a Riemann surface. This curvature corresponds to a distribution
consisting in a sum of $\delta-$functions concentrated at the branch points.
The classical equations of motions of the propagators $G(z,w)$ of the
fields $\varphi_k$ coming from the action \bosonizedaction\ are:
\eqn\electrostatics{\partial\bar \partial G(z;\alpha_i,\beta_j)=(1-2\lambda)
\delta^{(2)}(z,\infty)+\sum\limits_{i=1}^{N_\alpha}q_{k,\alpha_i}
\delta^{(2)(l)}(z,\alpha_i)+\sum\limits_{j=1}^{N_\beta}q_{k,\beta_j}
\delta^{(2)}(z,\beta_j)}
$G(z;\alpha_i,\beta_j)$ turns out to be
the Green function of electrostatics in the
presence of a charge $1-2\lambda$ at $z=\infty$ and fractional charges
$q_{k,\alpha_i}$ and $q_{k,\beta_j}$ at the branch points.
Only the $\delta-$function at the branch points $\alpha_i$ is
multivalued.
The total charge of the system is zero as it should be due to the
presence of the zero modes, as we will see in the next Section.
\vskip 1cm
\newsec{MULTIVALUED COMPLEX FIELDS ON THE PUNCTURED COMPLEX PLANE}
\vskip 1cm
In this Section we construct the Green function $K_\lambda(z,w)dz^\lambda
dw^{1-\lambda}$ of eq. \kl\ in terms of free fields using the
techniques of Section 2.
This tensor represents the two point function of the $b-c$ systems apart
from zero mode contributions and gives the vev's of the ghost currents
and of the energy momentum tensor as we have already seen.
We show in this way that $K_\lambda(z,w)dz^\lambda
dw^{1-\lambda}$ is a superposition of ratios of conformal blocks
of the kind \main.
Each conformal block correspond to a conformal field theory whose
monodromy properties are characterized by eqs. \localmonodromy\ and
\qkla-\qklb.\smallskip
First of all we consider the tensors $G_{\lambda,k}(z,w)$ that play the
role
of $G(z,w)$ in \main. For each
fixed value of $k$, they can be interpreted as the
propagators of the
sector of the $b-c$ fields having the same boundary conditions
at the branch points of $B_k(z)dz^\lambda$ and $C_k(z)dz^{1-\lambda}$:
\eqn\glk{G_{\lambda,k}^{(ll')}(z,w)dz^\lambda dw^{1-\lambda}={B_k^{(l)}(z)
C_k^{(l')}(w)\over z-w}dz^\lambda dw^{1-\lambda}}
Indeed, summing the partial propagators $G_{\lambda,k}
(z,w)$ of eq. \glk\ over $k$,
we get exactly the tensor $K_\lambda(z,w)$ which is,
as we have previously seen, the total propagator of the $b-c$
systems up to zero modes.
\smallskip
As in Section 2 we express
$G_{\lambda,k}^{(ll')}(z,w)$
in terms of free $b-c$ systems $\tilde b_k(z)dz^\lambda$ and
$\tilde c_k(z)dz^{1-\lambda}$ defined on the complex plane, $0\le k\le 2m-1$.
The effect of the branch points is simulated by the twist fields
$V_k(\alpha_i)$ and $V_k(\beta_j)$:
\eqn\glktf{
G_{\lambda,k}^{(ll')}(z,w)dz^\lambda dw^{1-\lambda}={{}_k<0|\tilde
b_k(z)
\tilde c_k(w)
({\rm z.m.})_k\prod\limits_{i=1}^{N_\alpha}V_k(\alpha_i)
\prod\limits_{j=1}^{N_\beta}V_k(\beta_j)|0>_k\over
{}_k<0|({\rm z.m.})_k\prod\limits_{i=1}^{N_\alpha}V_k(\alpha_i)
\prod\limits_{j=1}^{N_\beta}V_k(\beta_j)|0>_k}}
$|0>_k$ is the usual $SL(2,{\rm\bf C})$ invariant vacuum of the flat case and
$({\rm z.m.})_k$ represents an insertion of zero modes in eq. \glktf.
This will be necessary in order to set the total ghost charge to zero in the
correlators appearing in eq. \glktf.
To compute the number of zero modes we need to insert, we look
at the residues of
the ``current" $J_k(z)=\partial_z{\rm log}C_k^{(l)}(z)$.
{}From eqs. \ja-\jb\ we know already the total
ghost charge introduced by the presence of the branch points in
each  sector $k$ with independent
monodromy properties.
The computation of the total charge $q_{k,\infty}$
at infinity is easy to find and yields $q_{k,\infty}=1-2\lambda$.
All the $k$-sectors have the same charge at infinity and moreover
$q_{k,\infty}$
does not depend on $l$ confirming eq. \cpcurvature.
Summing all the charges at the branch points and at infinity we get
\eqn\zerocharge{\sum\limits_{l=0}^{2m-1}\left(\sum\limits_{i=1}^{N_\alpha}
\oint_{C_{\alpha_i}}+\sum\limits_{j=1}^{N_\beta}\oint_{C_{\beta_j}}+
\oint_{C_\infty}\right)dz\enskip\partial_z{\rm log}C_k^{(l)}(z)=N_{b_k}-
N_{c_k}}
where $C_{\alpha_i}$, $C_{\beta_j}$ and $C_\infty$ are closed infinitesimal
paths on the complex plane surrounding the points $\alpha_i$, $\beta_j$ and
$\infty$ respectively.
In eq. \zerocharge\ $N_{b_k}$ and $N_{c_k}$ are exactly the numbers of the
zero modes
$\Omega_{i_k,k}(z)dz^\lambda$ computed in the Appendix and having the
same behavior at the branch points of $B_k(z)dz^\lambda$ and $C_k(z)
dz^{1-\lambda}$.
In order  to get  nonvanishing amplitudes in eq. \glktf, we have therefore
to add the following insertion of zero modes:
\eqn\zmk{({\rm z.m.})_k=\prod\limits_{s=1}^{N_{b_k}}\tilde b_k(z_s)
\prod\limits_{t=1}^{N_{c_k}}\tilde c_k(z_t)}
Still we need the explicit expression of the twist fields.
These fields are derived in \ferdthree\ in the case of $D_3$ symmetric curves
using bosonization.
In the general case we just apply eq. \twistfields.
Let us introduce a set of
free scalar fields $\varphi_k(z)$ with propagator
$$<\varphi_k(z)\varphi_{k'}(w)>=-\delta_{kk'}{\rm log}(z-w)$$
Then the final form of the twist fields reads:
\eqn\vka{V_k^{(l)}(\alpha_i)={\rm exp}\left[i\oint_{C_{\alpha_i}}dt
\partial_t{\rm
log}\left[C_k^{(l)}(t)\right]\varphi_k(t)\right]}
\eqn\vkb{
V_k^{(l)}(\beta_j)={\rm exp}\left[i\oint_{C_{\beta_j}}dt \partial_t{\rm
log}\left[C_k^{(l)}(t)\right]\varphi_k(t)\right]}
Eq. \vkb\ can be further simplified and becomes:
\eqn\vkbe{V_k^{(l)}(\beta_j)=e^{-iq_{k,\beta_j}\varphi_k(\beta_j)}}
The asymptotic form of the twist fields $V_k(\alpha_i)$ at the branch points
is in agreement with eq. \ja\ and \localmonodromy.
Using the formulas given in Appendix B to compute the residues
at $\alpha_i$ and $\beta_j$ in eq. \vka\ we get in fact:
\eqn\vkae{V_k^{(l)}(\alpha_i)\sim e^{iq_{k,\alpha_i}(l)\varphi_k(\alpha_i)}
\qquad\qquad m\le l\le 2m-1}
\smallskip
Let us now show that the rhs of eq. \glktf\ is the desired subcorrelator
\glk\ following the formalism of Section 2.
Inserting in eq. \glktf\ the multivalued OPE's:
\eqn\ope{V_k
(\gamma)e^{-i\varphi_k(z)}e^{i\varphi_k(w)}={\rm exp}\oint_{C_\gamma}dt
\enskip\partial_t{\rm log}\left[C_k^{(t)}(t)\right]
{\rm log}\left({z-t\over w-t}\right)
:V_k(\gamma)e^{-i\varphi_k(z)}e^{i\varphi_k(w)}:}
with $\gamma=\alpha_i, \beta_j$ and remembering the contribution of the
charge at infinity, we have:
$$G_{\lambda,k}^{(ll')}(z,w)dz^\lambda dw^{1-\lambda}={dz^\lambda
dw^{1-\lambda}\over z-w}
\prod\limits_{s=1}^{N_{b_k}}
\left({z-z_s\over w-z_s}\right)\prod\limits_{t=1}^{N_{c_k}}\left(
{w-w_t\over z-w_t}\right)$$
\eqn\glka{
{\rm exp}\left [\left(
\sum\limits_{i=1}^{N_\alpha}
\oint_{C_{\alpha_i}}dt+\sum\limits_{j=1}^{N_\beta}\oint_{C_{\beta_j}}dt+
\oint_{C_\infty}dt\right)\partial_t{\rm log}\left[C_k^{(l)}(t)\right]
{\rm log}\left({z-t\over w-t}\right)\right]}
The contour $C=\sum_iC_{\alpha_i}+\sum_j C_{\beta_j}+{C_\infty}$
contains all the branch points as required in Section 2 and therefore we get:
the final result:
\eqn\glknew{G_{\lambda,k}^{(ll')}(z,w)dz^\lambda dw^{1-\lambda}=
{dz^\lambda dw^{1-\lambda}\over z-w}{C_k^{(l')}(w)\over
C_k^{(l)}(z)}\prod\limits_{i=1}^{N_{b_k}}\left({z-z_s\over w-z_s}\right)
\prod\limits_{t=1}^{N_{c_k}}\left({w-w_t\over z-w_t}\right)}
Remembering that $dz^\lambda/C_k^{(l)}(z)=B_k^{(l)}(z)
dz^\lambda$ from eq. \basislambda,
we conclude that eq. \glknew\ is the wanted solution of the Riemann
monodromy problem. The only difference from eq. \glk\
consists in the products involving the coordinates of the zero modes.
This is not a problem, since these terms coming from the zero modes
are rational functions of $z$ and do not modify the monodromy of the tensor
\glknew.\smallskip
Now we investigate the possibility of writing first order differential
equations for the Green function defined in eq. \glktf. In doing this we
regard the correlators \glktf\
as the correlators of a conformal field theory with
multivalued primary fields $V_k(\alpha_i)$.\smallskip
Since eq. \glktf\ is equivalent to eq. \glknew, we
need only to study the differential equations satisfied by the
$B_k(z)$'s for any value of $\lambda\in {\rm\bf Z}$.
It turns out that the functions $B_k(z)$ satisfy a differential equation
of the following kind:
\eqn\ptb{{dB_k^{(l)}(z)\over dz}=\sum\limits_{k'}A_{kk'}(z;
\alpha_i,\beta_j)B_{k'}^{(l)}(z)}
The elements of matrix $A_{kk'}(z;\alpha_i,\beta_j)$ are one forms
in $\Sigma_g$.
They are computed in Appendix B and we simply report that result:
\eqn\connection{A_{kk'}(z;\alpha_i,\beta_j)dz=\left(
mq_{k,\alpha_i}y^{-1}{dy\over dz}+q_{k,\beta_j}
\sum\limits_j{1\over z-\beta_j}\right)\delta_{kk'}dz}
The explicit dependence of $y^{-1}{dy\over dz}$ on $\alpha_i$ and $\beta_j$
can be found in eqs. (B.1-2).
Eq. \connection\ represents a one form with simple poles
at the branch points. Deriving the $B_k(z)$'s
with respect to $\alpha_i$ we get:
\eqn\ptbbranch{
\partial_{\alpha_i}B_k^{(l)}(z)=\tilde A_{kk'}(z;\alpha_i,\beta_j)
B_k^{(l)}(z)}
One can check that also $\tilde A_{kk'}(z;\alpha_i,\beta_j)$
is a one form in the variable $\alpha_i$ with simple poles
in $\alpha_i=z$ and $\alpha_i=\alpha_j$, $i\ne j$.
The residue at these points are exactly opposite to those of the matrix
$A_{kk'}(z;\alpha_i,\beta_j)$ and the two matrices differ only
by zero modes.
Analogous conclusions can be drawn deriving $B_k^{(l)}(z)$ with respect to the
branch points $\beta_j$. If we interprete eqs. \ptb\ as a parallel transport
\froa, \tk\ and \kohno,
then eqs. \connection\ and \ptbbranch\
provide the connection
in the variables $z$ and $\alpha_i$ respectively.
Using the above equations and the decomposition \splitting\ one can find
differential equations for all $n-$point functions of the $b-c$
systems on $\Sigma_g$.\smallskip
The matrices $A_{kk'}(z;\alpha_i,\beta_j)$ and
$\tilde A_{kk'}(z;\alpha_i,\beta_j)$ are not so simple as the usual
Knizhnik$-$Zamolodchikov equations. In fact
on a Riemann surface the two dimensional Poincar\'e group of world-sheet
symmetries is explicitly broken and the connections
$A_{kk'}(z;\alpha_i,\beta_j)$ and $\tilde A_{kk'}(z;\alpha_i,\beta_j)$
cannot be translational invariant as it happens in the flat case.
Therefore they have also a multivalued dependence on the variable $z$.
Eventually this is a consequence of eqs. \localmonodromy.
\smallskip
The twist fields \vka\ and \vkb\
represent particles with nonabelian braid group statistics inside the
amplitudes \glktf.
To conclude this section, we derive the exchange algebras of these operators.
The most difficult case occurs when two twist fields $V(\alpha_i)$
and $V(\alpha_{i'})$ are considered:
$$V^{(l_i)}_k(\alpha_i)V^{(l_{i'})}_k(\alpha_{i'})=$$
\eqn\vgvg{
{\rm exp}\left[-q_{k,\alpha_i}q_{k,\alpha_{i'}}\oint_{C_{\alpha_i}}ds
J_k^{(l_i)}(s)\oint_{C_{\alpha_{i'}}}ds'J_k^{(l_{i'})}(s'){\rm log}(s-s')
\right]V^{(l_{i'})}_k(\alpha_{i'})V^{(l_i)}_k(\alpha_i)}
To see how the twist fields are locally exchanged when $\alpha_i$ is
very near to $\alpha_j$, we have to compute two residues at the branch
points $\alpha_i$ and $\alpha_j$ in eq. \vgvg.
To do this it is sufficient to insert in the definition
of $J_k^{(l)}(z)$ given by eq. \jkdiff\ the form of $p(z)$
and $q(z)$ in terms
of the branch points provided by eqs. (B.1) and (B.2).
The remaining task
is a simple calculation of residues and the final result is:
\eqn\vava{V^{(l_i)}_k(\alpha_i)V^{(l_{i'})}_k(\alpha_{i'})=
e^{i\pi q_{k,\alpha_i}(l_i)q_{k,\alpha_{i'}}
(l_{i'})}V^{(l_{i'})}_k(\alpha_{i'})V^{(l_i)}_k(\alpha_i)}
\eqn\vavb{V^{(l_i)}_k(\alpha_i)V^{(l_j)}_k(\beta_{j})=
e^{i\pi q_{k,\alpha_i}(l_i)q_{k,\beta_j}(l_j)}
V^{(l_j)}_k(\beta_j)V^{(l_i)}_k(\alpha_i)}
\eqn\vbvb{V^{(l_j)}_k(\beta_j)V^{(l_{j'})}_k(\beta_{j'})=
e^{i\pi q_{k,\beta_j}(l_j)q_{k,\beta_{j'}}
(l_{j'})}V^{(l_{j'})}_k(\beta_{j'})V^{(l_j)}_k(\beta_j)}
where $q_{k,\alpha_i}(l_i)$ and $q_{k,\beta_j}(l_j)$
are defined in eqs. \qka\ and
\qkb.
\vskip 1cm
\newsec{CONCLUSIONS}
\vskip 1cm
One of the results obtained here is that the $b-c$ fields on an algebraic
curve
with $D_m$ group of symmetry can be decomposed into $2m$ sectors
propagating with different boundary conditions at the branch points.
Each $k$-sector, $0\le k\le 2m-1$, has a well defined propagator,
given by eq. \glktf\ and containing the multivalued twist fields
$V_k(\alpha_i)$.
The multivalued twist fields are primary fields and therefore to each
$k-$sector corresponds a multivalued conformal blocks.
We hope that with the formalism developed here one can treat also more
physical theories on algebraic curves than the $b-c$ systems. However,
the basic requirements are that the theory should be conformal and have
a lagrangian. This is not for example the case of theories based on free
scalar fields. The scalar fields, in fact, are not entirely conformal as
their propagator, with a logarithmic singularity, shows. As a
consequence an attempt to write an expansion of the kind \modeexpansion\
for the scalar fields is difficult, since they depend also on the
complex conjugate variable $\bar z$.
The free fermions, instead, are very interesting for superstring theory,
but unfortunately it is not so easy to treat the spin structures on
algebraic curves and therefore to construct analogues of the basis
\basislambda.\smallskip
Another result obtained is that we have shown
the presence of particles with nonstandard
statistics in the amplitudes of the $b-c$ systems and therefore of
string theory.
Following the procedure of Section 5 and using eq. \xcorr,
it is possible
to show that also the amplitudes of the free scalar fields contain multivalued
twist fields. The only problem is that there is no way to obtain an explicit
expression of these twist fields because
bosonization does not work in the case of the scalar fields.
\smallskip
It is natural to ask at this point if the twist
fields have some observable effect or if they are just an
artifact of our way of representing the Riemann surfaces as algebraic
curves.
First of all we remember that also in the case of the conformal field theories
there is a multivaluedness in the conformal blocks that disappears in the
physical correlation functions.
Despite of this fact, this multivaluedness is crucial in showing the
quantum group structure of conformal field theories.
In our case the multivaluedness on ${\rm\bf CP}_1$ of the amplitudes
is allowed and therefore also the presence of the twist fields.
The problem
is however complicated by the fact that the space-time geometry is not flat.
Surely a local observer, located in a system of reference in which the metric
on the Riemann surface is induced by the mapping $y(z):{\rm\bf CP}_1
\rightarrow\Sigma_g$, experiences the presence of the twist fields.
The existence of these operators is in fact proved in Section 5 using eqs.
\bclambda\ and \bcone, that represent the two point functions
obtained from the method of the fermionic construction of \vv.
Probably an observer in another system of reference would not confirm
the existence of the twist fields.
Unfortunately the calculations of the $n$-points functions in the case
of an arbitrary metric make use of the formalism of the theta functions
together with bosonization and the final results are not very explicit.
Therefore it is not easy to do a comparison of the observations
performed in the two different frames.
\vskip 1pt
The method presented here shows that a curved background can influence
the statistics inside the correlation functions of free field theories.
In Section 6 we have explained it in terms of electrostatics.
Finally we have realized in Section 7 examples of theories with
nontrivial braid group statistics \froa,\ref\mack{
G. Mack, V. Schomerus, {\it Nucl. Phys.} {\bf B370} (1992),
185.}.
It remains the problem to classify these theories.
To this purpose we only note that the
twist fields have nontrivial exchange relations but obviously they form
an associative algebra when more than two branch points
are permuted in the correlators of eq. \glktf. Therefore we can construct the
Yang-Baxter matrices corresponding to the exchange algebra \vava-\vbvb\
and look
if there are other integrable models yielding the same solutions of the
Yang-Baxter equations.
This has been done in ref. \ffbgs.
\vfill\eject
\centerline{APPENDIX A}
\vskip 1cm
The $2m$ branches of the solution of eq. \curve\ can be written as follows:
$$\cases{y^{(l)}(z)=e^{{2\pi i l\over m}}\enskip\root m \of{q(z)+ \sqrt{p(z)}}
\qquad 0\le l\le m-1\cr
y^{(l)}(z)=e^{{2\pi i l\over m}}\enskip\root m \of{q(z)- \sqrt{p(z)}}
\qquad m\le l\le 2m-5\cr}\eqno(A.1)$$
The branches are exchanged in the following way:
$$\eqalign{y^{(l)}(z)\rightarrow y^{(l+m)}(z)\qquad 0\le l\le 2m-1&\qquad
{\rm in }\quad \beta_1,\ldots,\beta_{N_\beta}\cr
(y^{(m)}(z),\ldots,y^{(2m-1)}(z))\rightarrow
(y^{(2m-1)}(z),y^{(m)}(z),\ldots,y^{(2m-2)}(z))&\qquad {\rm in}\quad
\alpha_1,\ldots,\alpha_{N_\alpha}\cr}\eqno(A.2)$$
We can rewrite eq. (A.2)
in a matrix form:
$$y^{(m)}(z)=(M_\gamma)_{m,l}y^{(l)}(z)\qquad \gamma=\alpha_i,
\beta_j\eqno(A.3)$$
The only nonvanishing elements of the monodromy matrices $M_{\beta_j}$
are $(M_{\beta_j})_{l+m,l}$. $M_{\alpha_i}$ has instead
the following block form: $M_{\alpha_i}={\rm diag}(I_m,S_m)$ where
$I_m$ is a $m\cdot m$ unit matrix and $S_m$ generates the $Z_m$ group of
permutations.
The monodromy matrices $M_{\alpha_i}$ and $M_{\beta_j}$ provide a
representation of the group $D_m$.\smallskip
The genus of the curve $\Sigma_g$ is given by the Riemann-Hurwitz
formula:
$$2g-2=2m((m-1)r-2)+2mr'\qquad\qquad mr\ge r'\eqno(A.4)$$
$$2g-2=2r'(m-1)+2mr'-4m\quad\qquad\qquad mr\le r'\eqno(A.5)$$
The behavior of a multivalued tensor near the branch points
$\alpha_i$ and $\beta_j$ is studied performing the following change of
variables:
$$t^m=z-\alpha_i\qquad\qquad\qquad{t'}^2=z-\beta_j\eqno(A.6)$$
$t$ amd $t'$ are the socalled local uniformizers in $\alpha_i$, $\beta_j$
respectively.
For example the behaviors of $y(z)$ and $p(z)$ at the branch points
is:
$$y^{(l)}(z)\sim\cases{{\rm const.}\qquad 0\le l\le m-1\cr
(z-\alpha_i)^{1\over m} \qquad m\le l\le 2m-1\cr}\qquad\qquad
\sqrt{p(z)}\sim(z-\beta_j)^{1\over 2}\qquad 0\le l\le 2m-1\eqno(A.7)$$
Using eq. (A.7) we are able to write the behaviors at the branch points
of the $\lambda$-differentials $B_k(z)dz^\lambda$.
\smallskip
It is possible to choose a basis of zero modes $\Omega_{i_k,k}(z)dz^\lambda$
in such a way that each element of the basis has the same monodromy properties
of $B_k(z)dz^\lambda$ given in eq. \basislambda:
$$\Omega_{i_k,k}(z)dz^\lambda=z^{i-1}B_k(z)dz^\lambda\qquad
1\le i_k \le N_{b_k}\eqno(A.8)$$
In order to determine the number of zero modes $N_{b_k}$ we can use the
following divisors.
We denote with $a^\nu_{(l)}$ a zero of multiplicity $\nu$ and with
$-a^\nu_{(l)}$ a pole of order $\nu$ occurring in the branch $l$ of a
meromorphic
$\lambda-$differential.
When $mr\le r'$ we have:
$${\rm div}[dz]=\sum\limits_{i=1}^{2mr}\alpha_i+
\sum\limits_{l=0}^{m-1}\sum\limits_{j=1}^{2r'}(\beta_j)_{(l)}-
\sum\limits_{i=0}^{2m-1}\infty^2_{(i)}\eqno(A.9a)$$
$${\rm div}[y(z)]=\sum\limits_{i=1}^{2mr}\alpha_i-
\sum\limits_{i=0}^{2m-1}\infty^r_{(i)}\eqno(A.9b)$$
When $mr< r'$ eqs. (A.9a) and (A.9b) become:
$${\rm div}[dz]=\sum\limits_{i=1}^{2r'}\alpha_i+
\sum\limits_{l=0}^{m-1}\sum\limits_{j=1}^{2r'}(\beta_j)_{(l)}-
\sum\limits_{i=0}^{2m-1}\infty^2_{(i)}\eqno(A.10a)$$
$${\rm div}[y(z)]=\sum\limits_{i=1}^{2r'}\alpha_i-
\sum\limits_{i=0}^{2m-1}\infty^{r\over m}_{(i)}\eqno(A.10b)$$
In order to eliminate possible branches at infinity,
$r'$ should be a multiple of $m$.
Studying the divisor of $\Omega_{i,k}(z)dz^\lambda$
it is also possible to prove that the total number of zero modes $N_{b_k}$
with the same behavior at the branch points of $B_k(z)dz^\lambda$
is given by (here we suppose $\lambda>0$):
$$N_{b_k}=1-2\lambda-\sum\limits_{l=0}^{2m-1}
\sum\limits_{i=1}^{N_\alpha}{1\over 2m}
q_{k,\alpha_i}(l)-\sum\limits_{j=1}^{N_\beta}q_{k,\beta_j}\eqno(A.11)$$
where $q_{k,\alpha_i}(l)$ is defined in eqs. \qkla\ and \qklb.
Summing over $k$ in eq. (A.11) and using eqs. (A.4)-(A.5) we get the total
number of zero modes
$N_b=(2\lambda-1)(g-1)$.
The $1-\lambda$ differentials have just a zero mode occurring
when  $\lambda=1$:
$$1=C_0(z)dz^{0}\eqno(A.12)$$
Therefore
$$N_{c_k}=\delta_{1\lambda}\eqno(A.13)$$
\vskip 1cm
\centerline{APPENDIX B}
\vskip 1cm
In this Appendix we prove that the basis \basislambda\
satisfies eq. \ptb.
We consider only the tensors $B_k(z)dz^\lambda$ because the proof for the
$1-\lambda$ differentials can be performed in a completely analogous way.
First of all from eq. \branchp\ we have:
$$q^2(z)-p(z)=\prod\limits_{i=1}^{N_\alpha}(z-\alpha_i)\qquad\qquad\qquad
p(z)=\prod\limits_{i=j}^{N_\beta}(z-\beta_j)\eqno(B.1)$$
Solving eq. (B.1) we get:
$$q(z)=\left(\prod\limits_{i=1}^{N_\alpha}(z-\alpha_i)+
\prod\limits_{j=1}^{N_{\beta_j}}(z-\beta_j)\right)^{1\over 2}\eqno(B.2)$$
Eq. (B.2) is useful because we can express in this way the polynomials
$q(z)$ and $p(z)$ appearing in eq. \curve\ in terms of the branch points.
Of course  not all the branch points $\alpha_i$  can be independent.
{}From eq. (B.1) they turn out to be functions of the other branch
points $\beta_j$ and of the zeros of $q(z)$.
This dependence of the $\alpha_i$ on the other branch points is necessary
because otherwise the polynomial $\prod\limits_{i=1}^{N_\alpha}(z-\alpha_i)+
\prod\limits_{j=1}^{N_{\beta_j}}(z-\beta_j)$ has not quadratic zeros.
This would be
in contradiction with the fact that, from eq. (B.2), this polynomial
should be equal to $q^2(z)$.
The fact that on the curves with nonabelian monodromy group the branch points
are not completely independent, makes it difficult to study the
properties of the twist fields and their OPE's under modular transformations
[26].
For the same reason it is not possible to compute explicitly
the matrix $\tilde A_{kk'}(z;\alpha_i,\beta_j)$ of eq. \ptbbranch\
apart from its pole structure.\smallskip
Now we consider the analytic tensor $B_k^{(l)}(z)dz^\lambda$ of
eq. (2.9) as a function $B_k^{(l)}(z)$ multiplied by the $\lambda$
differential $dz^\lambda$. For the functions $B_k^{(l)}(z)$ we compute the
ratio $(B_k^{(l)}(z))^{-1}(dB_k^{(l)}(z)/dz$.
The result is:
$$\left[B_k^{(l)}(z)\right]^{-1}{dB_k^{(l)}(z)\over dz}=mq_{k,\alpha_i}
y^{-1}{dy\over dz}+q_{k,\beta_j}\sum\limits_j{1\over z-\beta_j}\eqno(B.3)$$
More explicitely, from eq. \curve\ we have
$y(z)=(q+\sqrt{p})^{1\over m}$ and therefore
$$y^{-1}{dy\over dz}dz={1\over m}{dz\over q+\sqrt{p}}\left(q'+{1\over 2}
{p'\over \sqrt{p}}\right)\eqno(B.4)$$
The above differential has a simple pole in $z=\infty$ which does not
depend on the branches of $y(z)$.
Moreover, with the aid of the divisors of Appendix A,
it is clear that there are no
poles when $z\rightarrow \beta_j$ in eq. (B.4) despite of the fact that
${1\over 2}{p'\over \sqrt{p}}$ diverges in $z=\beta_j$.
However, in order to show that, we have to perform a change of coordinates
in eq. (B.4) switching to the local uniformizer \locunb.
The reason is that the differential $dz$ has exactly a zero in $\beta_j$
which cancels this singularity. Unfortunately the differential equation
\ptb\ is not covariant under transformations of coordinates
when $\lambda$-differentials ar involved. Motivated by these difficulties
in the approach of [12,18] we have introduced an alternative procedure
as explained in Section 2.
Finally, exploiting eqs. (B.1) and (B.2), it is possible
to see that
when $z\rightarrow\alpha_i$, the rhs of eq. \connection\ has only
a simple pole provided the branch $l$ of $y(z)$ is in the interval
$m\le l\le 2m-1$.
No other singularities are possible.
Therefore eq. (B.3) describes a RMP in which the 1-form matrix is
$$A_{kk'}(z)=mq_{k,\alpha_i}y^{-1}{dy\over dz}+q_{k,\beta_j}\sum\limits_j
{1\over z-\beta_j}\eqno\connection$$
\vskip 1cm
\centerline{ACKNOWLEDGEMENTS}
\vskip 1cm
The final version of this paper greatly benefitted from many fruitful
discussions with J. Sobczyk.
I am also grateful to Prof. J. Wess for the warm hospitality at Max Planck
Institute. This work was supported by a fellowship of the
Consiglio Nazionale delle Ricerche (CNR), Italy.
\vfill\eject
\newsec{FIGURE CAPTIONS}
\vskip 1cm
\item{Fig. 1} Lines of force on a wormhole in absence of charge.
The total effect of the lines is the appearance of virtual charges in the two
``mouths" of the wormhole.\smallskip
\listrefs
\bye